\documentclass{elsart}
\usepackage{psfig}
\usepackage{natbib}

\def\asca{{\it ASCA\ }}

\def\rosat{{\it ROSAT\/\ }}

\def\xmm{{\it XMM-Newton\ }}
\def\Hb{{$\rm H\beta$\ }}
\def\kms{{$\rm km\ s^{-1}$\ }}

\def\simgt{\lower 2pt \hbox{$\, \buildrel {\scriptstyle >}\over {\scriptstyle
\sim}\,$}}
\def\simlt{\lower 2pt \hbox{$\, \buildrel {\scriptstyle <}\over {\scriptstyle
\sim}\,$}}

\begin{document}

\runauthor{Boller}


\begin{frontmatter}

\title{\rosat Results on Narrow-Line Seyfert~1 Galaxies}

\author[MPE]{Th. Boller}
\address[MPE]{Max-Planck Institut f\"ur extraterrestrische Physik, Postfach
1312, 85741 Garching, Germany}
 
\begin{abstract}
The excellent soft X-ray sensitivity of the PSPC detector onboard the
\rosat satellite provided the first chance to study precisely the spectral
and timing properties of Narrow-Line Seyfert~1 galaxies. \rosat observations
of Narrow-Line Seyfert~1 galaxies have revealed 
(1) the existence of a giant soft X-ray excess, 
(2) a striking, clear correlation between the strength of the soft 
    X-ray excess emission and the FWHM of the \Hb line,
(3) the general absence of significant soft X-ray absorption by neutral hydrogen 
    above the Galactic column,
(4) short doubling time scales down to about 1000 seconds,
(5) the existence of persistent giant (above a factor of 10), and
    rapid (less than 1 day) X-ray variability in extragalactic sources.
The soft X-ray results on Narrow-Line Seyfert~1 galaxies indicate that
their black hole regions are directly visible, further supporting the
Seyfert~1 nature of these objects. 
The extreme X-ray properties of Narrow-Line Seyfert~1 galaxies make
them ideal objects for understanding many of the problems raised 
generally by
the Seyfert phenomenon. 
\end{abstract}


\end{frontmatter}


\section{Introduction}

This paper reviews the important \rosat contributions to the field of 
Narrow-Line Seyfert~1 (NLS1) research. The excellent soft X-ray 
sensitivity of the PSPC detector (Pfeffermann et al. 1987) onboard
the \rosat satellite (Tr\"umper 1983), the \rosat All-Sky Survey data (Voges 
et al. 1999), and deep \rosat PSPC and HRI pointed observations provided
the best opportunities to study NLS1s before the launches of \xmm and
{\it Chandra}. 
\rosat observations have triggered the rapid growth in the
definition of the phenomenological parameters of NLS1s throughout the
electromagnetic spectrum, as well as the theoretical modeling of their
exciting properties. This is clearly demonstrated with the papers presented
in these proceedings. Fifteen years after the definition of the peculiar optical 
properties of NLS1s (see the review article of Pogge), followed
by a period in which their importance had been suggested
(e.g. Halpern \& Oke 1987; Stephens 1989; Puchnarewicz et al. 1992),
NLS1s now represent an important class of the AGN family, holding many
keys to our understanding of the problems posed by the Seyfert 
phenomenon.


\section{The story of IRAS 13224--3809}
 
My NLS1 research started in 1992 with a \rosat PSPC AO-2 pointed 
observation of IRAS 13224--3809. The object was proposed for a
\rosat pointed observation because the \rosat All-Sky Survey observations
showed an unusually large X-ray luminosity for a Seyfert 2 classification,
as given at the time in the literature.
The remarkable X-ray luminosity of a few $\rm 10^{44}\ erg\ s^{-1}$ was
confirmed in the \rosat pointed observations, and, most surprisingly,
we found a light curve with unusually rapid and large-amplitude 
variability for a radio-quiet AGN. The shortest doubling time
was only 800 seconds, and the maximum variability amplitude was 
about a factor of 4 within a few hours. 
All this happened in mid-December 1992, and we thought it would be useful
to take an optical spectrum of this potentially important Seyfert galaxy.
I went to ESO and talked with Bob Fosbury on this matter. One of his
colleagues, Adline Caulet, was observing  during Christmas time
at the ESO 3.6~m telescope using the EFOSC spectrograph, and she actually 
took the first high-sensitivity
spectrum of IRAS 13224--3809 on December 24, 1992. The spectrum showed
a narrow \Hb line of less than about 1000 \kms FWHM and extremely strong
Fe II multiplet emission lines around 4500~\AA\ and 5200~\AA. Therefore,
the new optical observations revealed the NLS1 nature of IRAS 13224--3809,
and we found the first clear instance of strong and rapid X-ray variability
in a NLS1. During 1993--1994, Henner Fink and I worked
out the spectral properties of a few other NLS1 galaxies by comparing
the \rosat spectral slopes and the FWHM of \Hb for  NLS1s and broad-line
Seyfert~1 galaxies from the  Walter \& Fink (1993) sample.
We saw the first indications of an increasing spread of the
soft X-ray spectral slopes with decreasing line widths for Seyfert~1 
galaxies. The breakthrough in determining the soft X-ray properties
of NLS1s began in the middle of 1994. After I gave a talk at the Institute 
of Astronomy in Cambridge, U.K., Andy Fabian told me that
a student of his, Niel Brandt, was also working on the X-ray properties
of NLS1s (e.g. Akn~564) and had found similar results, and he asked me to 
talk with him. This was the beginning of a most constructive collaboration,
and within a couple of months we actually came up with our `Figure 8' showing 
the relation between the soft X-ray slope and the FWHM of the \Hb line
(Boller, Brandt \& Fink 1996). 

\section{NLS1s as an extreme of Seyfert activity}
NLS1s were found to exhibit extreme properties when compared with broad-line
Seyfert~1 galaxies. Many NLS1s show
\begin{itemize}
\item
A giant soft X-ray excess 
up to $\approx 1.5$~keV
\item
A larger than previously thought power-law slope diversity 
\item
Rapid and large-amplitude variability
\item
Indications of a highly ionized accretion disk
\item
Indications of a cooler accretion disk corona
\end{itemize}
\begin{figure}[htb]
\mbox{
\psfig{figure=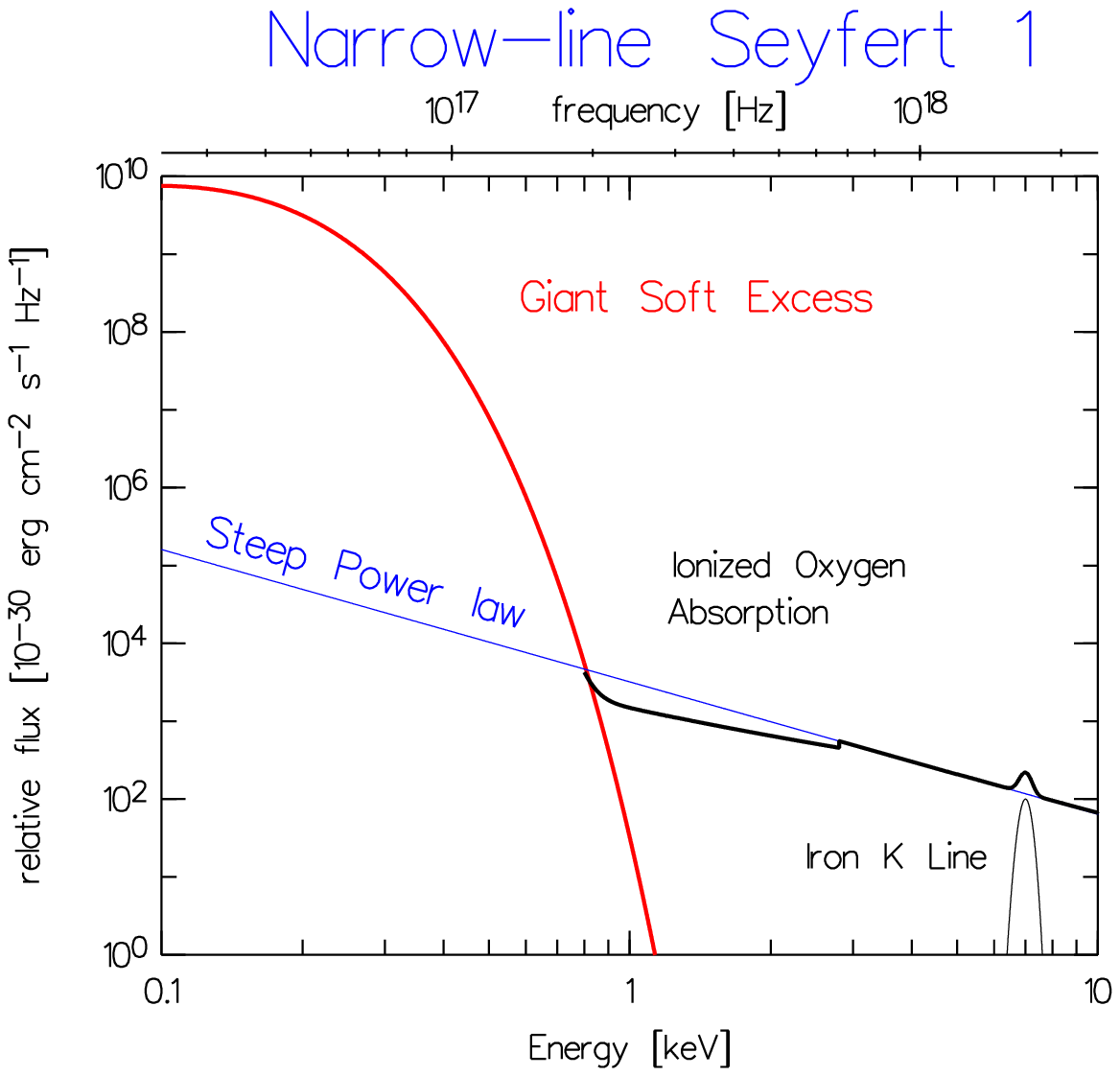,width=6.5cm}
\hspace{0.5cm}
\psfig{figure=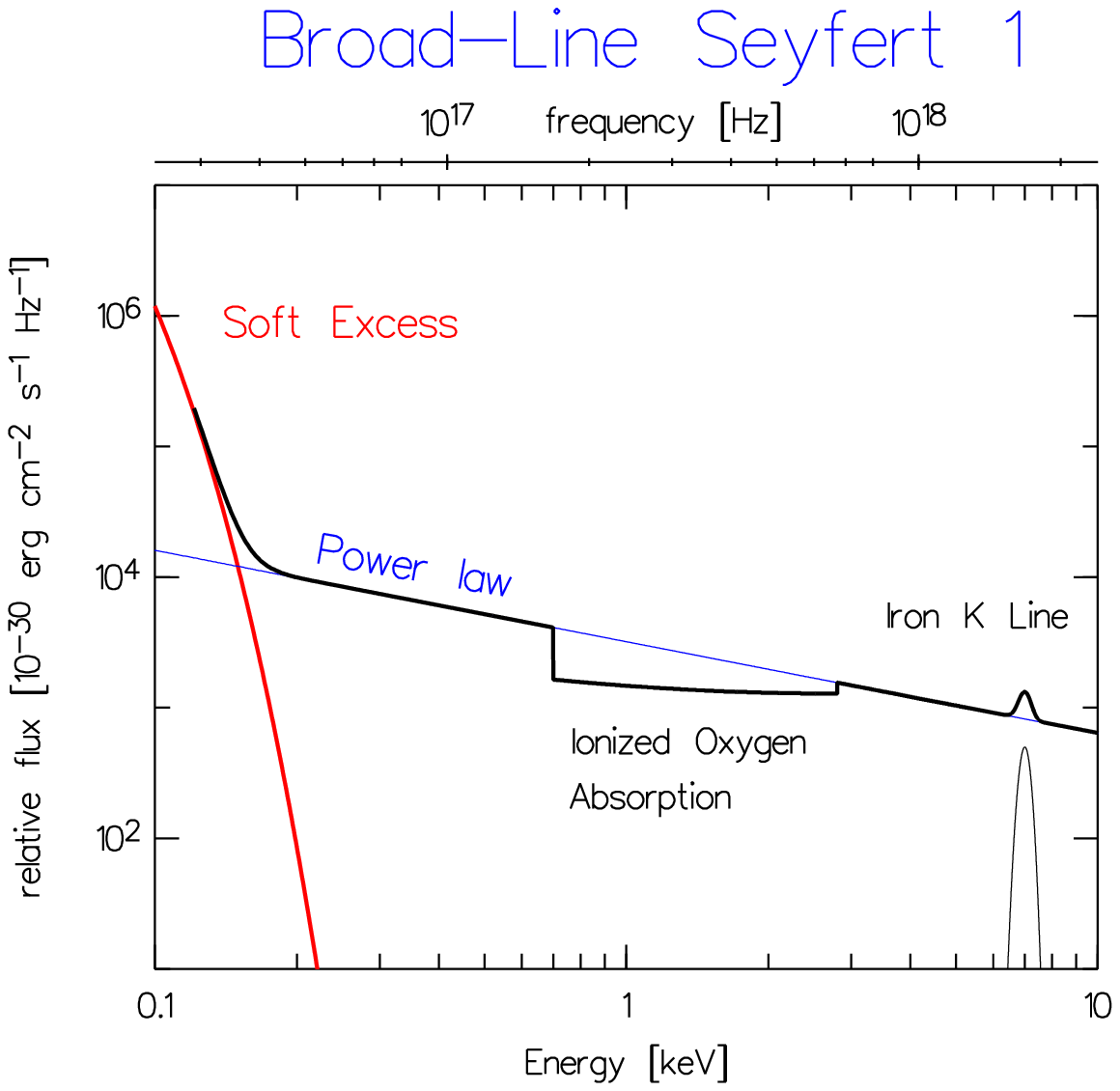,width=6.5cm}
}
\caption{
Schematic illustrations of the broad-band spectral energy distributions of
NLS1s (left panel) and broad-line Seyfert~1 galaxies (right panel).
The right panel is based on an idea of  A.C. Fabian. 
NLS1s often show extreme soft X-ray excess strengths and unusually
steep power-law X-ray continua.}
\end{figure}

Comparison of the schematic broad-band
X-ray energy distributions shown in Fig.~1 illustrates the extreme
soft X-ray excess and the generally steeper hard X-ray continua of NLS1s. 
The hard power-law slope diversity of NLS1s was discovered by 
Brandt, Mathur \& Elvis (1997).
Indications of an ionized Fe K$\alpha$ line are found in the NLS1 
Ton~S180 (Comastri et~al. 1998). 
In the case of a cooler accretion disk 
corona, steeper power-law slopes are expected due to the smaller
energy increase per scattering of an ultraviolet or soft X-ray photon.
A cooler accretion disk corona might
be due to stronger cooling by the huge soft X-ray excess photon
density. 
Accepted \xmm
observations of carefully selected NLS1s are expected to shed
further light on these issues. 
\newpage

\section{Optical emission-line correlations with X-ray continuum slopes}


\subsection{Correlations between the soft and hard X-ray continuum slopes
and the FWHM of \Hb} 

A strong relation between the \rosat photon index and the FWHM of the \Hb line 
has been detected in Seyfert~1 galaxies (e.g. Boller, Brandt \& Fink 1996; see 
the left panel of Fig.~2). 
While Seyfert~1 galaxies with \Hb FWHM greater than 3000 \kms have their
photon indices confined to a fairly narrow range,  
with a mean value of
about 2.3, NLS1s show an extremely large diversity in their values of the 
photon index,
reaching values of up to about 5. A region with values of the photon index above
about 3 and \Hb FWHM $\rm >$ 3000 \kms is not occupied by Seyfert~1 galaxies, and
it appears that Seyfert galaxies are not allowed to exist in this `zone of
avoidance.' In the \rosat energy band, the photon index serves as a measure
of the relative contributions of the soft X-ray excess emission and the
underlying power law. Sources with high values of photon
index are therefore thought to show the strongest soft excess emission. The
\Hb FWHM is a measure of the velocity dispersion in the Broad Line Region.
The soft X-ray emission probably arises within a few Schwarzschild radii 
of the central black hole in the accretion disk, and the  emission from the
Broad Line Region arises at significantly larger distances (light days to
light years; see the paper by Peterson in these proceedings). Therefore, the
correlation between these quantities suggests that the emission from the
accretion disk determines the velocity dispersion of the line emitting clouds
in the Broad Line Region. 

A similar correlation has been discovered by Brandt, Mathur \& Elvis (1997) between
the ASCA 2--10 keV photon index and the FWHM of $\rm H\beta$. NLS1s again show a larger
diversity in the distribution of their photon indices than broad-line Seyfert~1 galaxies.
In the 2--10 keV energy band, the spectral energy distribution is dominated by
the power-law contribution, which is thought to be primarily formed through Compton
scattering processes in the accretion disk corona. A complete understanding of the
steeper 2--10 keV X-ray continua in NLS1s has not  emerged up to now (see Section 3).
%

\begin{figure}[htp]
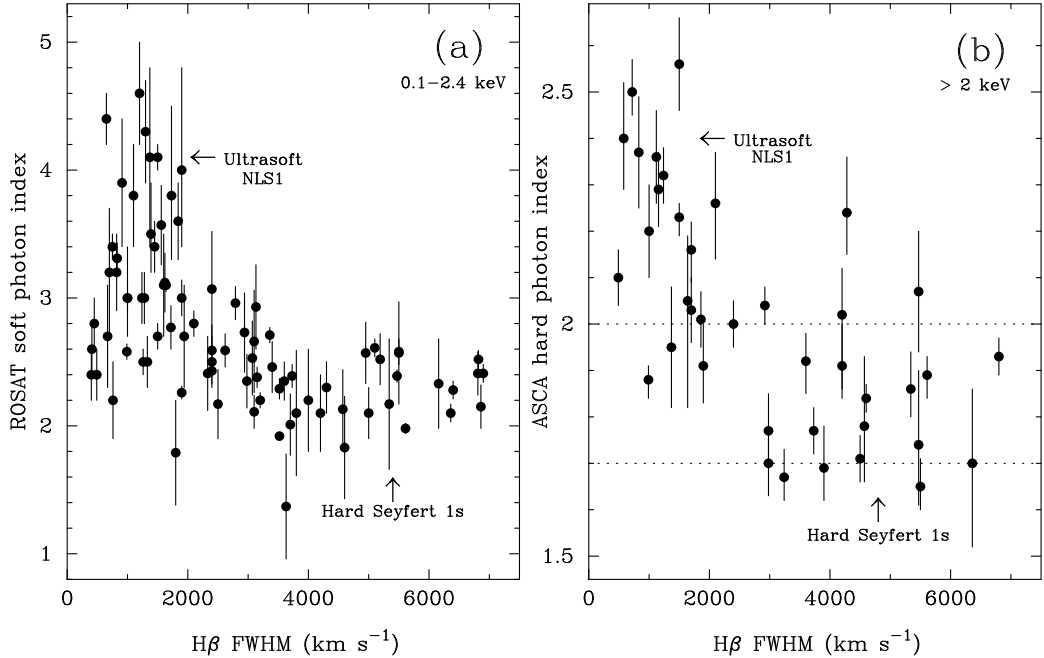

\mbox{
           \psfig{figure=boller_fig_3.ps,width=6.8cm,clip=}
           \psfig{figure=boller_fig_4.ps,width=6.8cm,clip=}}
         \caption[Anstieg des R\"ontgenkontinuums im weichen und harten
R\"ontgenbereich als Funktion der FWHM $\rm H\beta$-Linienbreite
]
{
{\bf Left Figure (a):} 
Photon index in the 0.1--2.4~keV energy range as a function of the FWHM 
of the $\rm H\beta$ line. The photon index serves as a measurement 
of the 
steepness of the X-ray continuum. All objects in the diagram are Seyfert~1
galaxies.  The NLS1s ($\rm H\beta$ line widths $\rm < 2000\ km\ s^{-1}$) are
taken from   Boller, Brandt \& Fink (1996). The broad-line Seyfert~1 galaxies
were investigated  by Walter \& Fink (1993). A significant correlation between
the slope of the X-ray  continuum and the FWHM of $\rm H\beta$ is
found. Seyfert~1 galaxies  with large line widths (FWHM $\rm H\beta > 2000\ km\
s^{-1}$) show a relatively
small dispersion in their values of the photon index with a mean of about 2.3. 
NLS1s show a strong dispersion in their values of the photon index. Some of these 
objects exhibit values of the photon index up to about 5.  
A region of the diagram, with values of the photon index larger than 3 and optical 
line widths larger than about 3000 $\rm km\ s^{-1}$, is not occupied by Seyfert~1 galaxies. 
This region is sometimes called the `zone of avoidance.'  
{\bf Right figure (b):}  
Photon index in the 2--10~keV energy range as a function of the FWHM of $\rm H\beta$ 
(Brandt, Mathur \& Elvis 1997). The measurements were obtained with the 
Japanese X-ray satellite {\it ASCA\/}. In the hard X-ray energy range, NLS1s show 
a stronger dispersion in their photon indices than previously thought.  
Typical values of the photon index in the hard X-ray energy band range between 
about 1.7 and 2.6. 
}
\end{figure}

\subsection{Seyfert~1 unification through physical processes}

A whole set of correlations between the X-ray spectral and timing properties
and optical emission-line properties have now been detected among Seyfert~1 galaxies.
The optical emission-line properties include the FWHM of H$\beta$, the strength
of the optical Fe~{\sc ii} emission (relative to H$\beta$) and the 
[O III]~$\lambda 5007$ strength.  
A compilation of X-ray and optical emission-line parameters is given in
Table 1. Plausible qualitative interpretations of the strong soft X-ray
excess observed in many NLS1s include a higher Eddington accretion rate and/or
lower black hole masses. Comptonization may play a role in explaining the
steep 2--10~keV power-law slopes found in NLS1s.
The postulated
higher ionizing photon density from a higher temperature accretion disk 
might
explain the  larger size of the Broad Line Regions in NLS1s (see the
paper of Peterson et al. in these proceedings).  
Boroson \& Green (1992) argued that a toroidal distribution of the 
Broad Line Region clouds
might prevent ionizing radiation from reaching the planar
Narrow Line Region (photon screening). This might cause  the weak [O III] 
emission seen 
in NLS1s. 
Relativistic Doppler boosting effects
or  X-ray flares above the accretion disk might explain the rapid
and giant amplitude variability detected in some NLS1s (see Section 5).

\begin{table}[htp]
\caption{Set of X-ray and optical correlations among Seyfert~1 galaxies (columns 1 and 3). 
Plausible underlying physical parameters are listed in column 2.}
\begin{tabular}{lcl}
\hline
{\bf Broad-line Seyfert~1}       & $ physical\ parameter$          & {\bf Narrow-line Seyfert~1} \\
Weak soft X-ray excess           & $\dot M, M$                     & Strong soft X-ray excess      \\
Flat power-law slope             & $Comptonization$                & Steep power-law slope      \\
Broad permitted lines            & $Ionizing\ photon\ density$     & Narrow permitted lines       \\
Weak Fe~{\sc ii}                 & $Cooler\ corona?$               & Strong Fe~{\sc ii}      \\
Strong [O III]                   & $Photon\ screening?$            & Weak [O III]      \\   
Moderate X-ray variability       & $Relativistic\ effects?$        & Extreme X-ray variability       \\
\hline
\end{tabular}
\end{table}

\section{The extreme X-ray variability of NLS1s}

In this section I concentrate on the most extreme cases of X-ray variability 
detected
in radio-quiet AGN, i.e. variability with an amplitude above 
a factor of about 10 in combination
with timescales of less than about one day. The X-ray light curves can be 
characterized
by strong flaring events interspersed with periods of relative quiescence.
The objects discussed below are the NLS1s IRAS~13224--3809, PHL~1092 and 
1ES~1927+654.  

\subsection{The persistent rapid and giant X-ray variability of IRAS 13224--3809}
The first systematic X-ray monitoring campaign for a NLS1 was 
performed in 1996 on 
IRAS~13224--3809. All relevant details are discussed in Boller et al. (1997). 
The most important observational facts can be summarized as: 
(1) the detection of multiple strong flaring events; 
(2) the most extreme variability over time is about a factor of 57
    in just two days;
(3) the light curve appears to be nonlinear or non-Gaussian in character;
(4) changes in the accretion rate or the ionization state of a warm 
    absorber
    are not considered plausible explanations for the extreme variability;
(5) the most probable explanations include strong Doppler boosting 
    effects
    arising  from temperature inhomogeneities in the accretion disk 
(see Sunyaev 1973; Guilbert, Fabian \& Rees 1993) or
partial obscuration by Compton thick matter (see the paper by Brandt \& Gallagher
in these proceedings). In Fig. 3 the \rosat HRI light curve of IRAS 13224--3809
obtained in 1996 is shown. 
\begin{figure}[htb]
\centerline{\psfig{figure=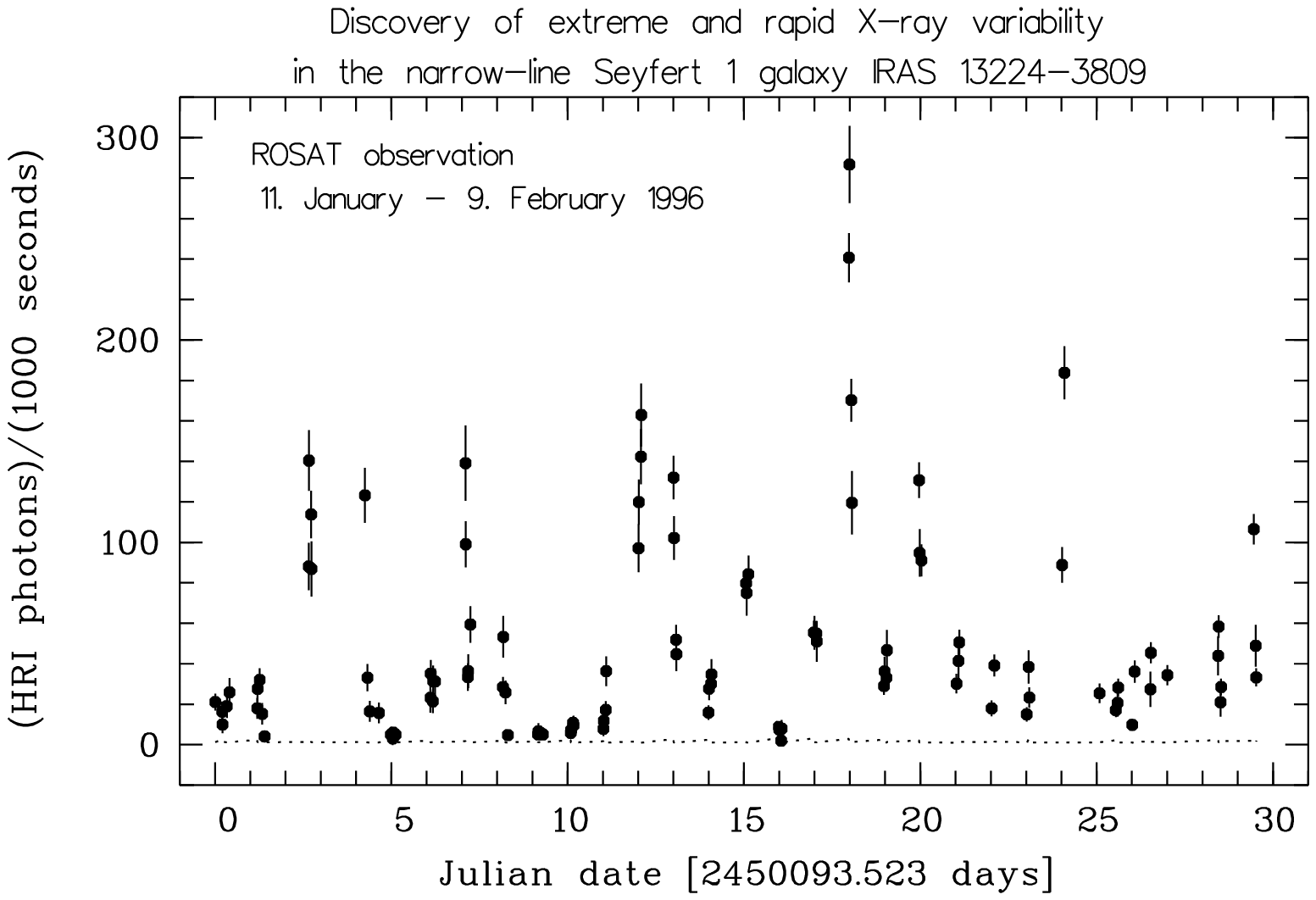,width=13cm,clip=}}
         \caption
{
\rosat HRI light curve of IRAS~13224$-$3809 obtained during a 30-day monitoring 
campaign between January 11, 1996 and February 9, 1996. 
The abscissa gives the Julian date minus 2450093.523 days. The dashed line shows 
the background count rate as a function of time. 
At least 5 giant-amplitude variations 
are clearly visible. 
IRAS 13224$-$3809 shows the most extreme X-ray variability 
measured thus far from active galactic nuclei.  
                        }
\end{figure}
\subsection{The large efficiency limit derived from the \rosat HRI light curve of
PHL~1092}
\begin{figure}[htb]
\centerline{\psfig{figure=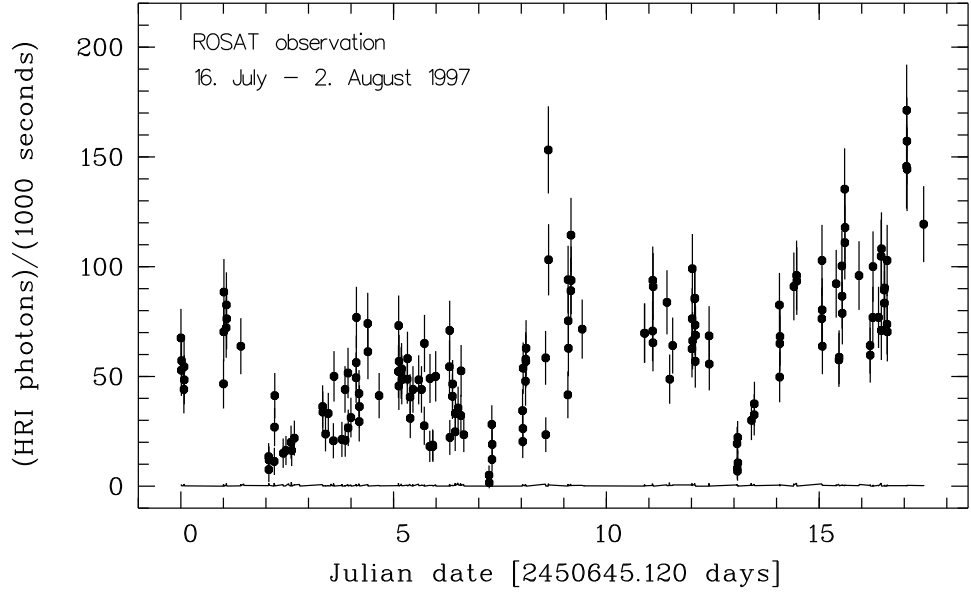,width=13cm,clip=}}
         \caption{
\rosat HRI light curve of PHL~1092 obtained during an 18-day observation 
between July 16, 1997 and August 2, 1997.
The abscissa gives 
the Julian date minus 2450645.120 days. The dashed line gives the 
background count rate as a function of time.
}
\end{figure}
In Figure 4 the \rosat HRI light curve of PHL 1092, a luminous narrow-line
quasar, is shown. The largest variability amplitude is about a factor
of 14 (see Brandt et al. 1999). 
The most extreme variability event has 
$\Delta L / \Delta t > 1.3 \cdot 10^{42}$~erg~s$^{-2}$, 
resulting in an extremely high efficiency of
$\eta \rm > 0.62\pm0.13$.
The \rosat HRI observation of PHL 1092 therefore further supports the
presence of relativistic motions of the X-ray emitting gas
causing giant and rapid flux variations in NLS1s.

\subsection{The persistent, rapid and giant X-ray variability of the X-ray
bright Einstein NLS1 1ES~1927+654}

With a mean 0.1--2.4~keV count rate of 
$\rm 1.26 \pm 0.026\ counts\ s^{-1}$,
1ES~1927+654 is one of brightest AGN detected in the \rosat All-Sky Survey.
Most interestingely, unusually large and persistent variability 
is detected, making this object, behind
IRAS~13224--3809 (Boller et al. 1997), the second radio-quiet
AGN showing this type of variability.
\begin{figure}[htb]
\centerline{\psfig{file=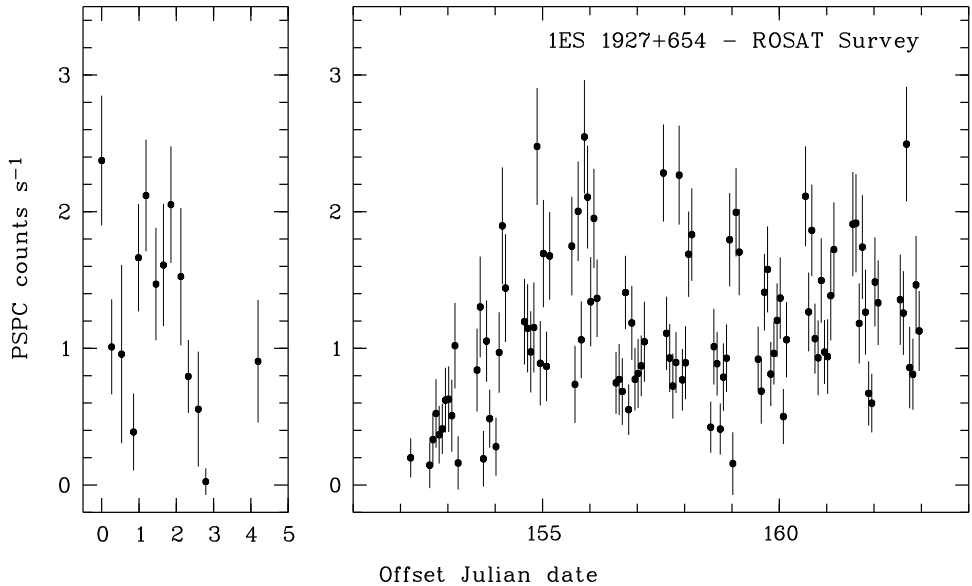,width=13cm,height=8cm,clip=}}
  \caption{
\rosat All-Sky Survey light curve of 1ES~1927+654 obtained 
between 1990 July 11--16
(left panel) and 1990 December 11--21  (right panel).
Persistent, rapid and giant X-ray variability is detected throughout the
\rosat PSPC observations. The maximum factor of variability is about
17.
     }
\end{figure}
1ES~1927+654 was observed during the `mini-survey' for about 5 days
between 1990 July 11 and 16 with
a total exposure time of 254 seconds.
During the normal survey scan operations, 1ES~1927+654
was observed for about 11 days
between
1990 December 11 and 21.
Unusually strong deviations from the mean count rate are detected
in the \rosat All-Sky Survey observations.
Figure 5 shows the \rosat PSPC light curve of 1ES~1927+654.
The left panel refers to the `mini-survey' observations in July 1990
and the right panel gives the count rate variations during the
survey scan observations in December 1990. 
Summing the six data points below a count rate of
0.2 $\rm counts\ s^{-1}$ results in a lower count rate level
of 0.146 $\rm counts\ s^{-1}$. Similarly, summing
the three data points above
2.4 $\rm counts\ s^{-1}$ results in an upper count rate level
of 2.51 $\rm counts\ s^{-1}$, the amplitude of  variability
is about 17.
Between days 154.017 and 154.150 the count rate increases
from 0.28 $\pm$ 0.212 to 1.90 $\pm$ 0.43
$\rm counts\ s^{-1}$, corresponding to a factor of about 8 within only
3.2 hours.

\begin{figure}[htb]
\centerline{\psfig{file=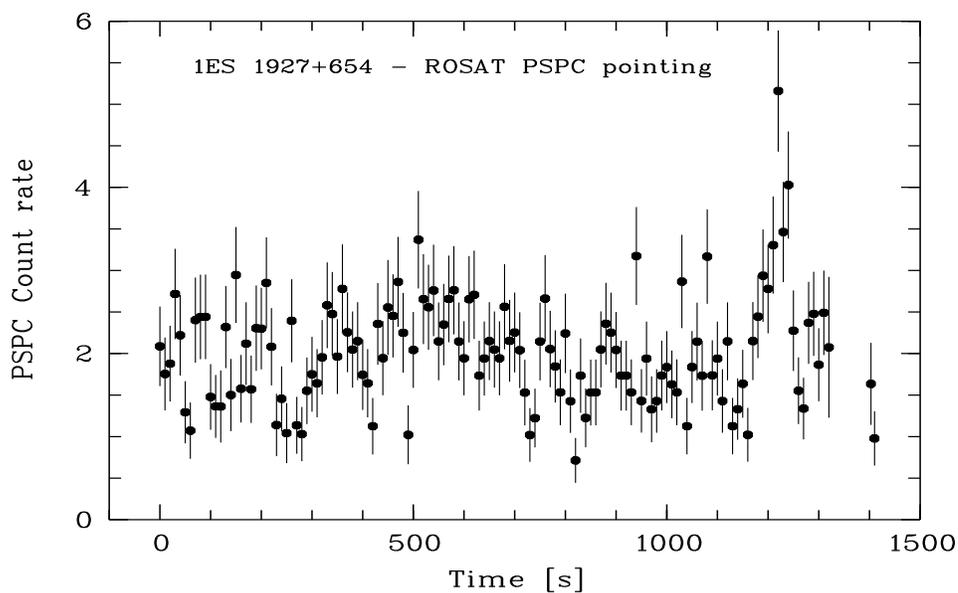,width=13cm,height=8cm,clip=}}
  \caption{\rosat PSPC light curve of 1ES~1927+654 obtained on
1998 December~8. A strong X-ray flare lasting for about 100 s
is detected near the end of the observations. The corresponding
isotropic energy is $\rm \approx 10^{46}\ erg$.
     }
\end{figure}

1ES~1927+654 was observed with the PSPC detector 
during the final observation
period of the \rosat satellite on December 8, 1998 (Figure 6).  
Using a bin size of 10 seconds, a strong X-ray flare becomes apparent
between 1110 and 1280 seconds after the beginning of the observations.
The count rate variations significantly exceed that of the variations
caused by the \rosat wobble. The isotropic energy emitted in this
strong X-ray flare is $E \rm = 1.7 \cdot 10^{46}\ erg$.
1ES~1927+654 is most probably another example showing the presence of 
relativistic motions of X-ray emitting gas in the nuclei of NLS1s.

\section{Summary}

Great progress in defining the observational properties of 
NLS1s has been achieved based on \rosat and \asca observations. 
The observations have shown many NLS1s to have characteristic, 
unique  and extreme X-ray properties. These include the strongest soft 
X-ray 
excess emission seen in Seyfert~1 galaxies, 
steep 2--10~keV power-law continua, and
extremely rapid and large-amplitude X-ray variability. 
The extreme properties of NLS1s have additionally stimulated the
theoretical modeling of many aspects of Seyfert activity.

\section{Acknowledgements}
The idea to organize a workshop on NLS1s began in June 1997 
at the `German-Japanese Workshop on Sky Surveys' in Potsdam, 
during a discussion with Niel Brandt and Martin Ward. In early 
1999 we finally decided to organize such a workshop. 
G\"unther Hasinger suggested to me that I  send an
application to the Wilhelm and Else Heraeus Stiftung to support the
workshop, which in May 1999 turned out to be successful.
Therefore, on behalf of the scientific organizing committee, I would like
to thank the WE-Heraeus Stiftung for their financial and organizational
support. I am especially grateful to Dr. E. Dreisigacker and Mrs. J. 
Lang for their constructive and effective help in preparing the workshop.
I would like to thank the Physikzentrum Bad Honnef, and especially Dr. 
DeBrus, for help finding accommodation outside the 
Physikzentrum and support during the workshop.
The scientific success of this joint MPE, AIP, ESO  workshop is 
due mainly to the work 
of the scientific organizing committee:
Jacqueline Bergeron (ESO), Niel Brandt (Penn State, USA), Suzy
Collin-Souffrin (France), Reinhard Genzel (MPE), Dirk Grupe (MPE), 
G\"unther Hasinger (AIP), Karen Leighly (Columbia, USA), 
Hagai Netzer (Tel Aviv, Israel), Joachim Tr\"umper
(MPE), Marie-Helene Ulrich (ESO) and Martin Ward (Leicester, UK).
On behalf of the scientific organizing committee, I would like to
thank all the participants for  
attending the workshop, for their input to the scientific discussions, 
and for submitting
their specific contributions, which we believe will be of great scientific 
interest for the astronomical community at large.

\vfill \eject


\begin{thebibliography}{999}




\bibitem{BBF96} Boller Th., Brandt W.N., Fink H., 1996, 
{\em Astronomy \& Astrophysics\/} {\bf 305} (1996) 53

\bibitem{BBF97} Boller Th., Brandt W.N., Fabian A.C., Fink H., 
{\em  Monthly Notices Royal Academic Society\/} {\bf 289} (1997) 393

\bibitem{BG} Boroson T.A. \& Green R.F.,
{\em The Astrophysical Journal Supplement Series} {\bf80} (1992)  109 

\bibitem{BME97} Brandt W.N., Mathur S., Elvis M., 
{\em  Monthly Notices Royal Academic Society\/} {\bf 285} (1997) 25

\bibitem{B99} Brandt W.N., Boller T., Fabian A.C., \& Ruszkowski M.,
{\em  Monthly Notices Royal Academic Society\/} {\bf 303} (1999) L53

\bibitem{C98} Comastri A., Fiore F., Guainazzi M., et al.,
{\em Astronomy \& Astrophysics} {\bf 333} (1998) 31

\bibitem{GFR83} Guilbert P.W., Fabian A.C., Rees M.J., 
{\em  Monthly Notices Royal Academic Society\/}  {\bf 205} (1983) 593

\bibitem{HO87}  Halpern J. \& Oke J.B., 
{\em The Astrophysical Journal} {\bf  312} (1987) 91

\bibitem{PBH87} Pfeffermann E., Briel U.G., Hippmann H., 
Kettenring G., Metzner G., Predehl P., Reger G., Stephan K.-H., Zombeck M.V.,
Chappell J., 
{\em Presented at the Society for Photo-Optical Instrumentation Engineers 
(SPIE), Berlin} (1987) 

\bibitem{P95}   Puchnarewicz E.M., Branduardi-Raymont G.,
Mason K. O. et al.
{\em  Monthly Notices Royal Academic Society\/} {\bf  276} (1995) 20

\bibitem{SS89}  Stephens, S., 
{\em Astronomical Journal} {\bf 97} (1989) 10

\bibitem{Sun73} Sunyaev R.A., 
{\em  Soviet Astronomy AJ} {\bf 16} (1973) 941

\bibitem{Tru83} Tr\"umper J., 
{\em Adv. Space Res.} (1983) {\bf 4}, 241

\bibitem{V1} Voges W., Aschenbach B., Boller Th. et al.,
{\em Astronomy \& Astrophysics} {\bf 349} (1999)  389


\bibitem{WF93} Walter R. \& Fink H.,
{\em Astronomy \& Astrophysics} {\bf 274} (1993) 105

\end{thebibliography}
\end{document}